\documentclass[
journal=jpccck, % for undefined journal
manuscript=article]{achemso}

\usepackage[version=3]{mhchem} % Formula subscripts using \ce{}

\newcommand*{\onlinecite}[1]{%
  \begingroup
    \romannumeral-`\x % remove space at the beginning of \setcitestyle
    \setcitestyle{numbers}%
    \cite{#1}%
  \endgroup   
}

\author{Jakub K. Sowa}
\email{jakub.sowa@oxon.org}

\author{Jan A. Mol}
\affiliation[Oxford]
{Department of Materials, University of Oxford, Oxford, OX1 3PH, UK}
\altaffiliation{School of Physics and Astronomy, Queen Mary University of London, London, E1 4NS, UK}
\author{Erik M. Gauger}
\affiliation[Heriot-Watt]
{SUPA, Institute of Photonics and Quantum Sciences, Heriot-Watt University, EH14 4AS, UK}

\title[\texttt{achemso} demonstration]
{Marcus Theory of Thermoelectricity in Molecular Junctions}

\begin{document}

\begin{abstract}
Thermoelectric energy conversion is perhaps the most promising of the potential applications of molecular electronics. 
Ultimately, it is desirable for this technology to operate at around room temperature, and it is therefore important to consider the role of dissipative effects in these conditions.
Here, we develop a theory of thermoelectricity which accounts for the vibrational coupling within the framework of Marcus theory.
We demonstrate that the inclusion of lifetime broadening is necessary in the theoretical description of this phenomenon. We further show that the Seebeck coefficient and the power factor decrease with increasing reorganisation energy, and identify the optimal operating conditions in the case of non-zero reorganisation energy.
Finally, with the aid of DFT calculations, we consider a prototypical fullerene-based molecular junction. We estimate the maximum power factor that can be obtained in this system, and confirm that C$_{60}$ is an excellent candidate for thermoelectric heat-to-energy conversion. 
This work provides general guidance that should be followed in order to achieve high-efficiency molecular thermoelectric materials.
\end{abstract}

The field of single-molecule electronics, first envisioned over 40 years ago, has been built on promises of delivering smaller, more efficient and cheaper electronic devices.\cite{aviram1974molecular}
Due to the significant technological advances in the field, molecular thermoelectric materials in particular have seen an upsurge in interest in recent years.\cite{reddy2007thermoelectricity,cui2017perspective,dubi2011colloquium,galperin2008inelastic}
It has been demonstrated, for instance, that quantum interference effects can be used to enhance the thermopower of single-molecule junctions.\cite{miao2018influence,bergfield2010giant,karlstrom2011increasing,finch2009giant} Furthermore, it has been shown how the molecular Seebeck coefficient can be tuned by varying the external environmental conditions.\cite{rincon2016molecular}
Particularly encouraging, however, are the recent groundbreaking studies which achieved electrostatic control of thermoelectric single-molecule junctions, and consequently relatively large thermoelectric power factors (which quantify the amount of energy that can be generated from a given temperature difference between the leads).\cite{kim2014electrostatic,gehring2017field}
Not only do these investigations open the door towards practical applications of molecular electronics but they also enable the exploration of the fundamental physics behind the thermoelectric energy conversion in nanoscale molecular devices.

The role of dissipative phenomena stemming from the electron-vibrational interactions in the molecular thermoelectric response is typically ignored when modelling experimental data.\cite{dubi2013possible,gehring2017field,kim2014electrostatic} However, as we will show here, these environmental interactions can have a tremendous effect on the molecular thermopower.
Our focus shall be the high-temperature thermoelectric behaviour of weakly coupled molecular junctions. 
In these conditions, one may expect to reach the regime of validity of Marcus description of the electron-vibrational interactions.\cite{marcus1956theory,marcus1985electron,nitzan2006chemical} Therein, the nuclear motion is treated classically, and the electron-vibrational coupling is accounted for by a single parameter: the reorganisation energy, $\lambda$ (see SI for further discussion).
Within this framework, the (near-)resonant charge transport is usually described by the Marcus-Hush-Chidsey theory\cite{chidsey1991free,zhang2008single,migliore2011nonlinear,bevan2015exploring,migliore2012relationship} which has been successfully used to account for experimental charge transport measurements on single-molecule junctions.\cite{jia2016covalently,migliore2013irreversibility,yuan2018transition}
Marcus-Hush-Chidsey theory treats the overall transport as a series of electron transfers occurring at the source and drain electrodes with the electron hopping on ($\gamma_l$) and off ($\bar{\gamma}_l$) rates given by (see for instance Ref.~\onlinecite{sowa2018beyond} for derivation):
\begin{align}\label{MT1}
& \gamma_l =  2\ \Gamma_l \int_{-\infty}^\infty \dfrac{\mathrm{d}\epsilon}{2\pi} f_l(\epsilon)  K_+(\epsilon) ~;\\   
& \bar{\gamma}_l = 2\ \Gamma_l \int_{-\infty}^\infty \dfrac{\mathrm{d}\epsilon}{2\pi} [1 - f_l(\epsilon)]  K_-(\epsilon) ~,\label{MT2}
\end{align}
where $f_l(\epsilon) = (\exp[(\epsilon-\mu_l)/k_B T_l]+1)^{-1}$, $\mu_l$ is the chemical potential, and $T_l$ is the electronic temperature  of the lead $l$. $\Gamma_l = 2\pi \lvert V_l\rvert^2 \varrho_l$ where $V_l$ is the molecule-lead coupling strength, and $\varrho_l$ the constant density of states in the lead $l$ (wide-band approximation). 
The energy dependent hopping rates are:
\begin{equation}\label{KMHC}
    K_\pm(\epsilon) = \sqrt{\dfrac{\pi}{4\lambda k_{\mathrm{B}} T_{\mathrm{ph}}}} \exp\left( -\dfrac{[\lambda \mp (\epsilon - \bar{\varepsilon}_0)]^2}{4\lambda k_{\mathrm{B}} T_{\mathrm{ph}}}\right)  ~. 
\end{equation}
In the above, $\lambda$ is the Marcus reorganisation energy, and $T_{\mathrm{ph}}$ is the temperature of the phononic (vibrational) environment of the molecular energy level in question.

As discussed above, the position of the molecular level, $\bar{\varepsilon}_0$, can be controlled by applying the gate potential $V_\mathrm{g}$ via $\bar{\varepsilon}_0 = - \lvert e \rvert V_\mathrm{g}$, where for simplicity we have set the zero-gate position of the molecular level to zero (and taken the lever-arm of 1).

In this Letter, we address the following questions: (i) Can Marcus-Hush-Chidsey theory be used to describe the thermoelectric effect in molecular junctions and, if not, how can this be remedied? (ii) What is the optimal parameter regime, in terms of reorganisation energy and lifetime broadening, for an efficient thermoelectric energy conversion? (iii) What power factors can be reached with realistic molecular systems?

%\section{Theory}
To address the above questions we study a model molecular junction, and assume that the molecular structure possesses a single  electronic energy level close to the Fermi energy of the unbiased leads.
The overall system is found in a non-equilibrium steady-state owing to slow electronic relaxation. The electron population on the molecular system can be easily determined using the quantum master equation yielding the current:\cite{sowa2018beyond}  
\begin{equation} \label{current}
    I = \dfrac{e}{\hbar} \ \dfrac{\gamma_\mathrm{L} \bar{\gamma}_\mathrm{R} - \gamma_\mathrm{R} \bar{\gamma}_\mathrm{L}}{\gamma_\mathrm{L} + \gamma_\mathrm{R} + \bar{\gamma}_\mathrm{L} +\bar{\gamma}_\mathrm{R}} ~.
\end{equation}
Each of the leads has its own temperature $T_l$ where $l = \mathrm{L}, \mathrm{R}$. For convenience, we set $T_\mathrm{R} = T$ and $T_\mathrm{L} = T + \Delta T$.
In this work, we are interested in the linear response regime where $ \Delta T \equiv (T_\mathrm{L}-T_\mathrm{R})  \rightarrow 0 $. 
While, in principle, the phononic temperature $T_{\mathrm{ph}}$ can be different than the electronic temperature of the leads, here we set $T_{\mathrm{ph}} = T$ (which is consistent with the limit $ \Delta T \rightarrow 0 $).
% Revision:
Furthermore, as usual for the Marcus description of electron transfer, we assume that the vibrational environment is found in thermal equilibrium at all times. Once again, this assumption is consistent with the limit of vanishing $\Delta T$. We note, however, that non-equilibrium vibrational effects may play an important role for large $\Delta T$, and the above assumption therefore constitutes a limitation of our theoretical approach (and Marcus-type approaches in general).

The temperature difference between the two electrodes induces a thermal current, the sign and magnitude of which depends on the position of the molecular energy level ($\bar{\varepsilon}_0$). 
Under open-circuit conditions, a thermal voltage $V_\mathrm{th}$ is established such as to nullify the thermal current flowing through the junction.
The main quantity of interest is the Seebeck coefficient which is defined, in the linear response regime and for vanishing $\Delta T$, in terms of the open-circuit voltage drop across the junction as:
\begin{equation}
    S = - \left. \lim_{\Delta T \rightarrow 0} \dfrac{V_{\mathrm{th}}}{\Delta T} \right|_{I = 0} ~.
\end{equation}
Following the seminal work of Beenakker and Staring,\cite{beenakker1992theory} we note that in the linear regime the current can be expanded as:
\begin{equation}\label{Beenaker}
    I = G V + G_{\mathrm{th}} \Delta T +...
\end{equation}
and therefore the Seebeck coefficient is given by: $S = G_{\mathrm{th}}/G$.
$G$ and $G_{\mathrm{th}}$ can be obtained by expanding the rates $\gamma_\mathrm{L}$ and $\bar{\gamma}_\mathrm{L}$ to the first order in $V$ and $\Delta T$: $\gamma_\mathrm{L} = \gamma_\mathrm{L}^{(0)} + \gamma_\mathrm{L}^{(V)} V + \gamma_\mathrm{L}^{(\Delta T)} \Delta T$, (and equivalently for $\bar{\gamma}_\mathrm{L}$)
and inserting them into Eq.~\eqref{current}.
We obtain:
\begin{align}
    G_{\mathrm{th}} &= \dfrac{\gamma_\mathrm{L}^{(\Delta T)} \bar{\gamma}_\mathrm{R} - \bar{\gamma}_\mathrm{L}^{(\Delta T)} \gamma_\mathrm{R} }{\gamma_\mathrm{L}^{(0)} + \gamma_\mathrm{R} + \bar{\gamma}_\mathrm{L}^{(0)} + \bar{\gamma}_\mathrm{R}} ~; \\
    G &= \dfrac{\gamma_\mathrm{L}^{(V)} \bar{\gamma}_\mathrm{R} - \bar{\gamma}_\mathrm{L}^{(V)} \gamma_\mathrm{R} }{\gamma_\mathrm{L}^{(0)} + \gamma_\mathrm{R} + \bar{\gamma}_\mathrm{L}^{(0)} + \bar{\gamma}_\mathrm{R}} ~,
\end{align}
and consequently
\begin{equation}\label{seebeck}
    S = \dfrac{\gamma_\mathrm{L}^{(\Delta T)} \bar{\gamma}_\mathrm{R} - \bar{\gamma}_\mathrm{L}^{(\Delta T)}  \gamma_\mathrm{R}}{\gamma_\mathrm{L}^{(V)} \bar{\gamma}_\mathrm{R} - \bar{\gamma}_\mathrm{L}^{(V)} \gamma_\mathrm{R} }~.
\end{equation}
Finally,
\begin{align}
\gamma_l^{(\Delta T)} &= 2 \Gamma_l \int_{-\infty}^\infty \dfrac{\mathrm{d} \epsilon}{2\pi} f_l'(\epsilon) \left( -\dfrac{\epsilon}{T}\right) K_+(\epsilon) \\
\bar{\gamma}_l^{(\Delta T)} &= 2 \Gamma_l \int_{-\infty}^\infty \dfrac{\mathrm{d} \epsilon}{2\pi} f_l'(\epsilon) \left(\dfrac{\epsilon}{T}\right) K_-(\epsilon) \\
\gamma_l^{(V)} &= 2 \Gamma_l \int_{-\infty}^\infty \dfrac{\mathrm{d} \epsilon}{2\pi} f_l'(\epsilon) \left(-K_+(\epsilon) \right)\\
\bar{\gamma}_l^{(V)} &= 2 \Gamma_l \int_{-\infty}^\infty \dfrac{\mathrm{d} \epsilon}{2\pi} f_l'(\epsilon) \left(K_-(\epsilon) \right)~.
\end{align}

The alternative to using the above expressions is numerically finding the potential difference $V$ which nullifies the current in Eq.~\eqref{current} for a given $\Delta T$ (and confirming the linear relationship between $V$ and $\Delta T$ as $\Delta T \rightarrow 0$).
Both of these methods yield identical values of $S$.

%\section{Results}
\begin{figure}
    \centering
    \includegraphics{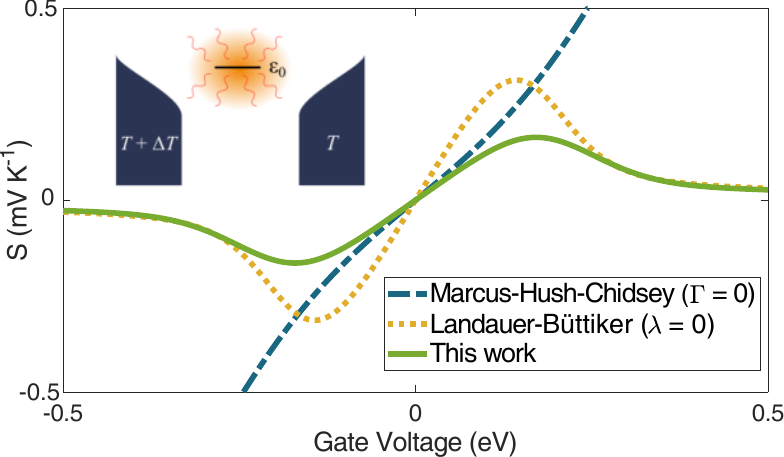}
    \caption{Seebeck coefficient as a function of the gate voltage for $\lambda = 0.1$ eV and $\Gamma = 5$ meV at $T = 300$ K obtained using the Marcus-Hush-Chidsey theory, Landauer-B\"uttiker approach, and the method introduced here. Inset: a schematic of the system studied here.}
    \label{fig1}
\end{figure}
We begin by evaluating the Seebeck coefficient using Marcus-Hush-Chidsey (MHC) theory, as well as Landauer-B\"uttiker (LB) approach which does not account for the electron-vibrational coupling. We consider a symmetrically coupled energy level, $\Gamma_\mathrm{L} = \Gamma_\mathrm{R} = 5$ meV, with the reorganisation energy $\lambda = 100$ meV.
As can be seen in Fig.~\ref{fig1}, in agreement with previous attempts,\cite{craven2017electron} MHC theory yields an unphysical description of the thermoelectric effect. The Seebeck coefficient increases almost linearly as the position of the molecular energy level is shifted away from the resonance reaching virtually limitless values for large $V_\mathrm{g}$. Furthermore, as it can be readily inferred from Eq.~\eqref{seebeck}, the Seebeck coefficient is independent of the strength of the molecule-lead coupling.
Both of these shortcomings  are characteristic of approaches perturbative in  molecule-lead interactions, and stem from the lack of lifetime broadening in these descriptions.\cite{koch2004thermopower,craven2017electron}

As we have recently shown,\cite{sowa2018beyond} lifetime broadening can be incorporated in the Marcus description of charge transport by replacing the rates in Eq.~\eqref{KMHC} with:
\begin{equation}\label{KH}
    K_\pm(\epsilon) = \mathrm{Re} \bigg[\sqrt{\dfrac{\pi}{4\lambda k_{\mathrm{B}} T_{\mathrm{ph}} }} \exp\left( \dfrac{(\Gamma - \mathrm{i}\nu_\pm)^2}{4\lambda k_{\mathrm{B}} T_{\mathrm{ph}}}\right) \times \mathrm{erfc}\left(\dfrac{\Gamma - \mathrm{i}\nu_\pm}{\sqrt{4\lambda k_{\mathrm{B}} T_{\mathrm{ph}}}}\right)\bigg] ~,
\end{equation}
where $\nu_\pm = \lambda \mp(\epsilon - \varepsilon_0 )$, $\Gamma = (\Gamma_\mathrm{L} + \Gamma_{\mathrm{R}})/2$ is the lifetime broadening, and $\mathrm{erfc}(x)$ denotes the complementary error function.
Notably, as we have previously discussed, by setting $\Gamma = 0$ one trivially recovers MHC theory, while setting $\lambda \rightarrow 0$ together with Eq.~\eqref{current} yields the Landauer-B\"uttiker approach for a single non-interacting level.\cite{sowa2018beyond} 
Our approach recovers the physical behaviour of the Seebeck coefficient, see Fig.~\ref{fig1}. Near the resonance it is in an agreement with Marcus-Hush-Chidsey theory, and slowly converges to the LB approach at large gate voltages.
% Revision:
This approach can be straightforwardly generalised to go beyond the Marcus-type description of the vibrational environment. This can be done simply by replacing the energy-dependent rates $K_\pm (\epsilon)$ with expressions from Ref.~\onlinecite{sowa2018beyond}, see SI for details. Doing so, however, requires the detailed knowledge of phononic spectral density. 
In the SI we further benchmark the Marcus approach used here against its generalisation. We find that, as expected, the two methods are in agreement in the case of coupling to relatively low-frequency modes, i.e.~when the high-temperature assumption of Marcus theory is justified.

\begin{figure*}
    \centering
    \includegraphics{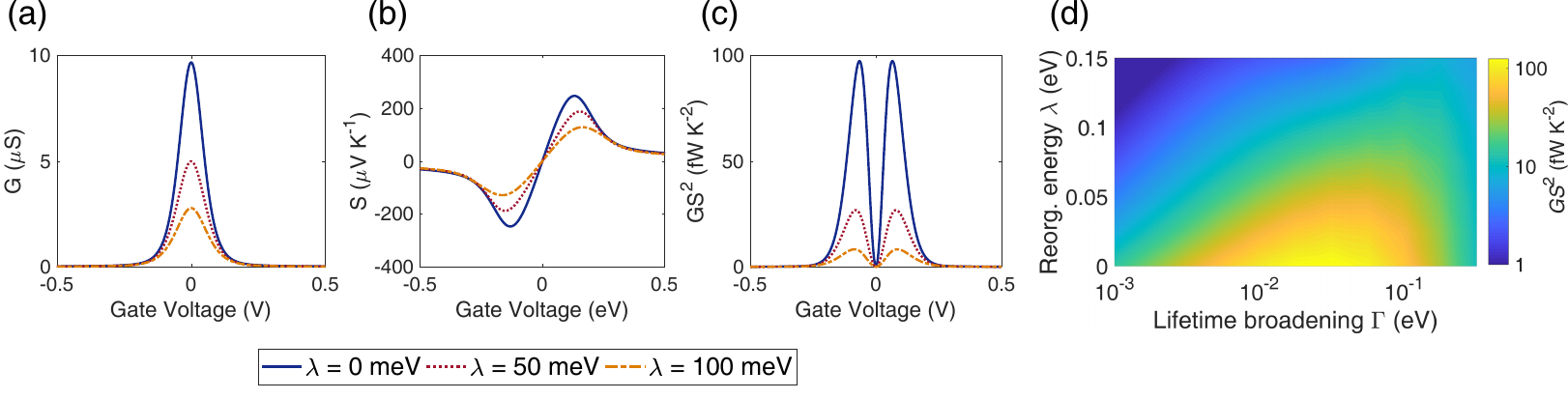}
    \caption{(a) Zero-bias conductance $G$, (b) Seebeck coefficient $S$, and (c) Power factor $GS^2$ as a function of the gate voltage with $\Gamma = 10$ meV, and for various values of the reorganisation energy, $\lambda$. (d) Maximum power factor as a function of lifetime broadening $\Gamma$, and the reorganisation energy $\lambda$. $T = 300$ K throughout.}
    \label{fig2}
\end{figure*}

We move to the second point raised at the beginning of this work.
In Fig.~\ref{fig2}, we plot the zero-bias conductance $G$, the Seebeck coefficient $S$, and the power factor $GS^2$ for different values of the reorganisation energy. Both the conductance and the Seebeck coefficient decrease with increasing $\lambda$, which can be understood as follows: The zero-bias conductance generally decreases with increasing strength of the electron-vibrational coupling due to reduced Franck-Condon overlap for the ground-state-to-ground-state transition.\cite{koch2005franck} This is also true within the Marcus picture.
The trend of the Seebeck coefficient can, on the other hand, be explained by the fact that the reorganisation energy acts as a source of broadening of the energy-dependent hopping rates in Eq.~\eqref{KH} (analogously to what is qualitatively predicted by the Mott formula \cite{lunde2005mott}).
As a consequence of the above trends, the power factor  decays very rapidly with $\lambda$, Fig.~\ref{fig2}(c).
As we demonstrate in the SI, qualitatively similar behaviour can be obtained when modelling the environmental interactions using a dephasing approach within the non-equilibrium Green function formalism.\cite{bihary2005dephasing,penazzi2016self,cresti2006electronic} 
In practical realisations of single-molecule thermoelectric materials one should, therefore, strive to minimise the environmental vibrational coupling.
Examples of systems attractive in this context (i.e.~with low reorganisation energy) include the C$_{60}$ fullerene (\textit{vide infra}), as well as certain polyaromatic hydrocarbons such as pentacene or graphene nanoribbons.\cite{sancho2009charge}
Notably, molecular structures with low reorganisation energy are also of fundamental interest in optoelectronic materials and organic field-effect transistors.

Given that a certain reorganisation energy is intrinsic to a considered system what value of lifetime broadening $\Gamma$ is optimal for an efficient heat-to-energy conversion? 
Generally speaking, the molecular conductance increases with $\Gamma$, while the Seebeck coefficient follows the opposite trend. Consequently, the power factor reaches a maximum for a certain value of $\Gamma$.
It is well known that, in the absence of electron-vibrational coupling and for a single-level molecular system, the optimum is achieved for $\Gamma \sim 1.1 k_\mathrm{B} T$.\cite{gehring2017field} 
In Fig.~\ref{fig2}(d) we plot the maximum power factor that can be achieved for given $\lambda$ and $\Gamma$. As can be seen in Fig.~\ref{fig2}, the electron-vibrational coupling has a more detrimental effect on the zero-bias conductance than on the Seebeck coefficient. Consequently, for a non-zero reorganisation energy, the power factor $GS^2$ reaches the maximum at larger values of $\Gamma$ as compared to the $\lambda =0$ case.
The electron-vibrational interactions should, therefore, be taken into account when designing the optimal junction geometry.

\begin{figure*}
    \centering
    \includegraphics{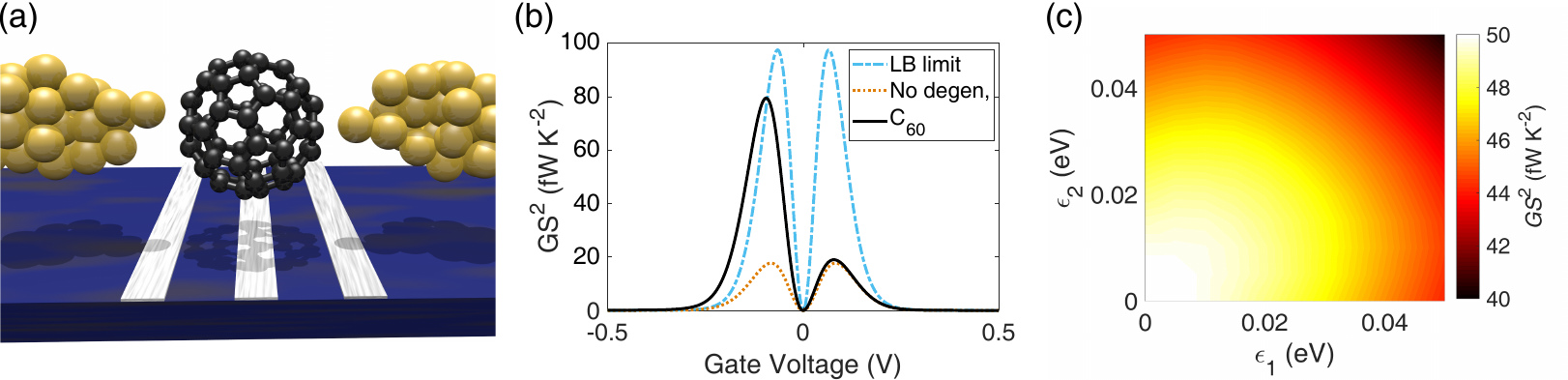}
    \caption{(a) Schematic illustration of a C$_{60}$-based molecular junction. (b) The power factor $GS^2$ as a function of the gate voltage calculated for: Landauer-B\"uttiker limit ($n=1$, $\lambda = 0$); no electronic degeneracy ($n=1$, $\lambda = 67$ meV); C$_{60}$-based molecular junction ($n=6$, $\lambda = 67$ meV). $\Gamma = 10$ meV. (c) The power factor calculated for the C$_{60}$-based molecular junction in the presence of symmetry breaking at $\Gamma = 5$ meV. $\Delta \varepsilon_1$ and $\Delta \varepsilon_2$ are the energies of the excited LUMO levels above the ground-state of the $N+1$ charge state. $T=300$ K throughout. }
    \label{fig3}
\end{figure*}

We next estimate the maximum power factor that can be obtained for a prototypical C$_{60}$ fullerene-based junction,\cite{evangeli2013engineering,kim2014electrostatic} schematically shown in Fig.~\ref{fig3}(a). The relevant value of $\lambda$ is half of the reorganisation energy for the corresponding self-exchange reaction:\cite{nelsen1987estimation}
\begin{equation}
    \lambda = \dfrac{1}{2} \left[ E_-(Q_0) - E_-(Q_-) + E_0(Q_-) - E_0(Q_0) \right]~,
\end{equation}
where $E_-(Q_-)$ is the total energy of C$_{60}^-$ in its equilibrium geometry, and $E_-(Q_0)$ is the total energy of the negatively charged C$_{60}$ in the equilibrium geometry of the neutral species, and equivalently for $E_0(Q_0)$ and $E_0(Q_-)$.
The calculation in Gaussian 09 with B3LYP functional and the 6-31G(d,p) basis set\cite{frisch2009gaussian} yields the value of the reorganisation energy of $\lambda = 67$ meV.

Furthermore, departing from conventional approaches, we note that the LUMO level of the C$_{60}$ fullerene is three-fold spatially-degenerate, and thus has the overall degeneracy of $n=6$.
This means that for the charge transition studied herein (between a singly-degenerate ground-state of $N$ charge state and six-fold degenerate ground-state of $N+1$ charge state), the electron hopping onto the molecule is six times as likely as it would have been in the absence of this degeneracy.
As we show in the SI, the Seebeck coefficient is unaffected by the degeneracy of the level involved in the charge transport. On the other hand, the conductance increases with increasing $n$, and the position of the conductance peak shifts towards the more negative gate voltage, see SI.\cite{beenakker1991theory}
Due to these two effects, the power factor increases as the degeneracy of the considered level is increased, see Fig.~\ref{fig3}(b). 
Highly symmetric molecular structures, such as C$_{60}$, are therefore very attractive candidates for an efficient thermoelectric heat-to-energy conversion. 
Furthermore, in Fig.~S4 in the SI we repeat the calculation from Fig.~\ref{fig2}(d) and plot the maximum power factor that can be obtained for the fullerene-based junction (accounting for both the reorganisation energy and the six-fold degeneracy of the LUMO level) as a function of the lifetime broadening $\Gamma$. We estimate that, at $T=300$ K, the maximum value of the power factor that can be achieved for this system is $(GS^2)_{\mathrm{max}} \approx 120$ fW K$^{-2}$ (although we note that deviations from the Marcus description of environmental interactions can result in more efficient thermoelectric energy conversion, see SI). For comparison, within the Landauer-B\"uttiker treatment the maximum power factor  (for a singly-degenerate electronic level not coupled to a vibrational environment) is $(GS^2)_{\mathrm{max}} \approx 135$ fW K$^{-2}$.
The high degeneracy of the LUMO level of the C$_{60}$ molecule can therefore, at least partially, offset the detrimental effects of the environmental interactions, as can also be seen in  Fig.~\ref{fig3}(b). 
We note however that populating the spatially degenerate electronic states can give rise to Jahn-Teller (JT) distortion.\cite{sowa2018spiro}
The effect of JT distortion on the thermoelectric energy conversion is beyond the scope of this work -- this issue will be addressed in the future.
%Revision:
Furthermore, in the SI, we study the deviation of the thermal voltage from the linear-response behaviour considered here. We show that only relatively small deviations of $V_\mathrm{th}$ from the linear response can be observed for experimentally-relevant temperature differences between the leads.

Deposition of a fullerene molecule into a junction, or its chemical functionalisation,\cite{gehring2017field,lau2015redox} can result in some degree of symmetry breaking. This will split the otherwise degenerate LUMO levels, and can be expected to decrease the conductance of the junction, and thus have an adverse effect on the power factor.
In Fig.~\ref{fig3}(c) we show the maximum power factor that can be obtained for the fullerene-based junction in the presence of symmetry breaking for relatively weak molecule-lead coupling: $\Gamma = 5$ meV. 
Therein, the otherwise triply spatially-degenerate states of the $N+1$ charge state are split with the two higher energy levels lying $\Delta \varepsilon_1$ and $\Delta \varepsilon_2$ above the ground-state.
We assume that the reorganisation energy remains unchanged in the presence of this symmetry breaking, and assume a lack of interference between these three pathways, see SI for details of the calculation.
As shown in Fig.~\ref{fig3}(c), we find that the performance of the studied molecular junction is largely unaffected by the symmetry breaking as long as the energy splittings are smaller than the thermal (or in principle also the lifetime) broadening. As the result, the performance of a fullerene-based thermoelectric junction is robust to relatively small symmetry breaking (for instance one that may occur due to non-covalent interactions with the metallic electrodes) at $T=300$ K.
On the other hand, the chemical functionalisation of the fullerene core (of the type performed in experimental studies of Refs.~\onlinecite{gehring2017field,lau2015redox}) can induce considerable splitting of the otherwise triply-degenerate LUMO level, see SI. Such structures are therefore significantly less attractive candidates for an efficient thermoelectric energy conversion.

In summary, we have shown that due to the absence of lifetime broadening Marcus-Hush-Chidsey theory cannot be used to describe the thermoelectric effect in molecular junctions. 
Instead, we have formulated an intuitive and relatively simple approach, still in the spirit of Marcus theory, which allows us to capture the effects of electron-vibrational coupling in the thermoelectric heat-to-energy conversion. 
Notably, in contrast to the Landauer-B\"uttiker formalism, our approach further allows for an easy inclusion of the degeneracy of the relevant energy levels.

Using our novel approach, we have demonstrated that the vibrational coupling (non-zero $\lambda$) has a detrimental effect on both the electronic conductance and the Seebeck coefficient. 
Consequently, we have shown that molecular systems with small reorganisation energy and of high symmetry, such as the C$_{60}$ fullerene, should be used to achieve high thermoelectric power factors.
Our work further suggests that one should strive to minimise the environmental (outer-sphere) vibrational coupling by resorting to an appropriate device geometry. Furthermore, the electron-vibrational interactions also need to be considered when optimising the molecule-lead coupling strengths. 
Finally, our work provides a general framework for describing the behaviour of high-temperature molecular thermoelectric materials, and should prove valuable in the future development of efficient molecular technologies.

%\section*{Author contributions} JKS conceived the study, derived analytical results, and performed the calculations. JKS wrote the manuscript with input from EMG. All authors discussed the approach and results, and commented on the final manuscript.

\suppinfo
Details of the theoretical methods, additional results, and further discussion.

\acknowledgement
The authors thank Pascal Gehring for useful discussions.
J.K.S. thanks the Clarendon Fund, Hertford College and EPSRC for financial support. E.M.G. acknowledges funding from the Royal Society of Edinburgh and the Scottish Government, J.A.M. acknowledges funding from the Royal Academy of Engineering.
The authors would like to acknowledge the use of the University of Oxford Advanced Research Computing (ARC) facility in carrying out this work. http://dx.doi.org/10.5281/zenodo.22558

\providecommand{\latin}[1]{#1}
\makeatletter
\providecommand{\doi}
  {\begingroup\let\do\@makeother\dospecials
  \catcode`\{=1 \catcode`\}=2 \doi@aux}
\providecommand{\doi@aux}[1]{\endgroup\texttt{#1}}
\makeatother
\providecommand*\mcitethebibliography{\thebibliography}
\csname @ifundefined\endcsname{endmcitethebibliography}
  {\let\endmcitethebibliography\endthebibliography}{}

\end{document}